\documentclass[
 reprint,
superscriptaddress,
prl,
 amsmath,amssymb,
 aps,
]{revtex4-1}

\usepackage{graphicx}
\usepackage{dcolumn}
\usepackage{bm}
\usepackage[colorlinks= true, linkcolor=blue, citecolor=blue, urlcolor=blue]{hyperref}
\usepackage{lipsum}
\usepackage{soul}
\usepackage{bbold}
\usepackage[normalem]{ulem}
\usepackage{mathtools}
\usepackage{xcolor}
\usepackage{ragged2e}
\usepackage{mathtools}
\usepackage{amsmath} 
\usepackage{mathtools}
\usepackage{amsmath}
\usepackage{braket}
\usepackage{float}
\usepackage{lipsum}  
\usepackage{svg}
\let\vec\mathbf
\renewcommand\thesection{\arabic{section}}
\usepackage[sort&compress]{natbib}
\usepackage{mhchem}
\usepackage{changepage}
\usepackage{float}
\usepackage{mathrsfs}
\usepackage{tikz}
\usepackage{braket}

\begin{document}

\title{Inference from gated first-passage times}

\author{Aanjaneya Kumar}
\email{kumar.aanjaneya@students.iiserpune.ac.in}
\affiliation{Department of Physics, Indian Institute of Science Education and Research, Dr. Homi Bhabha Road, Pune 411008, India.}
\author{Yuval Scher}
\email{yuvalscher@mail.tau.ac.il}
\affiliation{School of Chemistry, Center for the Physics \& Chemistry of Living Systems, Ratner Institute for Single Molecule Chemistry, and the Sackler Center for Computational Molecular \& Materials Science, Tel Aviv University, 6997801, Tel Aviv, Israel}
\author{Shlomi Reuveni}
\email{shlomire@tauex.tau.ac.il}
\affiliation{School of Chemistry, Center for the Physics \& Chemistry of Living Systems, Ratner Institute for Single Molecule Chemistry, and the Sackler Center for Computational Molecular \& Materials Science, Tel Aviv University, 6997801, Tel Aviv, Israel}
\author{M. S. Santhanam}
\email{santh@iiserpune.ac.in}
\affiliation{Department of Physics, Indian Institute of Science Education and Research, Dr. Homi Bhabha Road, Pune 411008, India.}

\date{\today}

\begin{abstract}
First-passage times provide invaluable insight into fundamental properties of stochastic processes. Yet, various forms of \emph{gating} mask first-passage times and differentiate them from actual \emph{detection times}. For instance, imperfect conditions may intermittently gate our ability to observe a system of interest, such that exact first-passage instances might be missed. In other cases, e.g., certain chemical reactions, direct observation of the molecules involved is virtually impossible, but the reaction event itself can be detected. However, this instance need not coincide with the first collision time since some molecular encounters are infertile and hence gated. Motivated by the challenge posed by such real-life situations we develop a universal---model-free---framework for the inference of first-passage times from the detection times of gated first-passage processes. In addition, when the underlying laws of motions are known, our framework also provides a way to infer physically meaningful parameters, e.g. diffusion coefficients. Finally, we show how to infer the gating rates themselves via the hitherto overlooked \emph{short-time regime} of the measured detection times. The robustness of our approach and its insensitivity to underlying details are illustrated in several settings of physical relevance.
\end{abstract}

\maketitle

\emph{Introduction.---}The importance of first-passage processes is recognized universally across scientific disciplines, owing to their ubiquity and wide-ranging applications \cite{redner2001,GreHolMet,metzler_first-passage_2014,weiss1967first,chicheportiche2014some,iyerbiswas_first-passage_2016,zhang_first-passage_2016}. How long does it take for a chemical reaction to be triggered? Or what is the time taken for an order to be executed in the stock market? These disparate examples fall under the purview of first-passage processes, where the \emph{first-passage time} is now established as an indispensable tool to quantify the time taken for a given task to be completed.

In several practically relevant scenarios, however, the completion of a task also relies on additional constraints. For example, for a chemical reaction to be triggered, two reactants must collide. Additionally, the collision must be fertile, i.e., the reactants must be in a reactive internal state during collision. This internal state acts like a ``\emph{gate}": a reaction can only happen when the gate is ``open", i.e. the molecules are in their reactive internal state. The macroscopic kinetics of these so called \emph{gated} reactions has a history spanning over four decades now \cite{mccammon1981gated,szabo1980first,szabo_stochastically_1982,northrup1982rate,weiss1986overview,whitmore_first-passage-time_1986,zhou1996theory,berezhkovskii1997smoluchowski,makhnovskii_stochastic_1998,benichou2000kinetics,bandyopadhyay2000theoretical,bressloff2015stochastically,gopich2016reversible}, and more recently, the study of single-particle gated reactions has gained interest \cite{budde1995transient,spouge1996single,bressloff2015escape1,bressloff2015escape2,godec_first_2017,mercado-vasquez_first_2021,bressloff2020diffusive,toste2022arrival,mercado2019first,scher_unified_2021,scher_gated_2021}. 

\begin{figure}
\justifying
\includegraphics[width=0.95\linewidth]{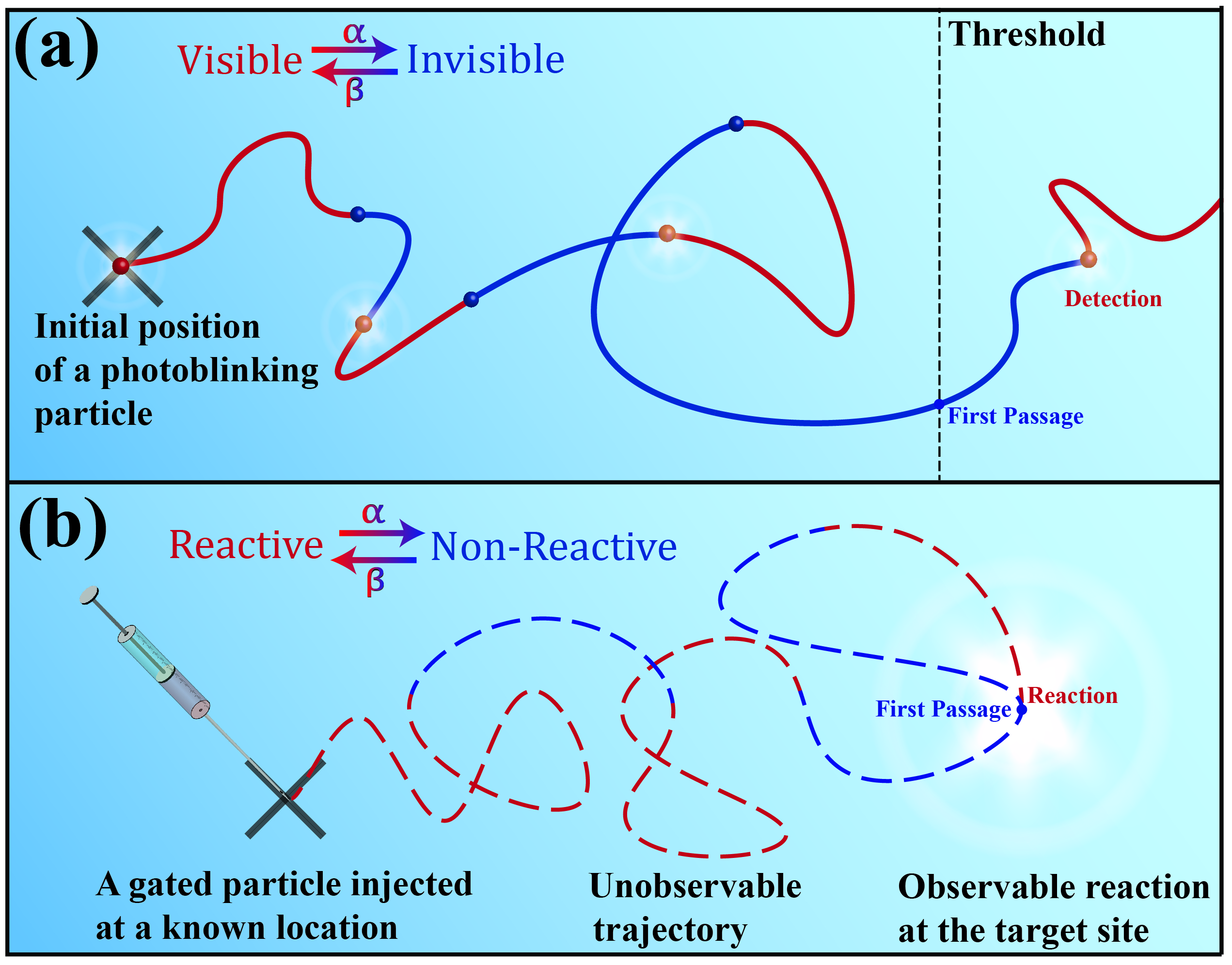}
\caption{
 Instances highlighting the need for inference in gated first-passage processes. (a) Detection of threshold crossing under intermittent sensing. Consider single particle tracking of a photoblinking particle. The first-passage properties of the particle can be mischaracterized as the particle can cross the threshold while being in its invisible state. (b) Gated chemical reaction or target search. Imagine a situation where tracking of the particle is not possible, and the only observable is the reaction time. For such processes, we show how the first-passage time distribution, and other relevant observables, can be inferred from detection times.}
\label{fig:schematic}
\end{figure}

While the terminology of `gating' is unique to  reaction kinetics, numerous examples fall under the wide umbrella of gated processes. An important one is that of intermittently observed stochastic time-series, where the underlying cause for intermittent observations can include energy costs of continuous observations, imperfect detection conditions, or simply, a faulty sensor  \cite{wsn-rev-1, wsn-rev-2,mobileSensor,HsinLiu,intsen-robot}. Irrespective of the reasons behind such intermittent observations, an important consequence is that key features of the time-series can be missed. In particular, in the increasingly relevant field of extreme and record statistics of time-series, a crucial quantity is the time taken to cross a specific threshold for the first time. However, intermittent detection of the time-series can lead to a gross mischaracterization of the statistics of such events \cite{zarfaty2021discrete,zarfaty2022discrete,kumar_first_2021,makarov2022effect,song2023effect}. In such cases, the relevant quantity is the \emph{first detection time}, which denotes the first time the time-series is \emph{observed} above the threshold \cite{kumar_first_2021}.

Figure~\ref{fig:schematic} exemplifies two instances where gating arises naturally: (a) Single-particle tracking of an intermittently observed particle, which transitions between a visible state and an invisible state. For example, a wide class of fluorophores undergo photoblinking \cite{ha1997quantum,lu1997single,dickson1997off,peterman1999fluorescence,bout1997discrete,nirmal1996fluorescence,kuno2001off,schuster2005blinking,claytor2009accurately,khatua2009micrometer}. Other reasons for such gating can be the intermittent loss of focus on a moving particle in 3-dimensions \cite{hansen2018robust} or slow frame acquisition rate \cite{kues2002single}; (b) A gated chemical reaction or target search, where tracking of the particle is not possible, and the reaction time is the only measurable quantity. Such instances may arise in cellular signalling driven by narrow escape \cite{holcman_escape1,holcman_escape2} and among fluorescent probes \cite{giepmans2006fluorescent}. 

In both examples illustrated in Fig.~\ref{fig:schematic}, the first-passage time statistics carry invaluable information, but are inaccessible to direct measurement. In such scenarios, a crucial challenge is to reliably \emph{infer}  these statistics and other fundamental properties of interest.

In this Letter, we address this challenge and solve it.  First, we show how the first-passage time density can be inferred from gated observations via a model-free formalism, which upon specification of the underlying laws of motions can be further used to infer physically meaningful  parameters (e.g., the diffusion coefficient). Second, using the joint knowledge of the gated (observed) and ungated (inferred) first-passage time densities, we establish that the overlooked \emph{short-time regime} of the gated detection time distribution can be leveraged to obtain the gating rates.

\emph{Modeling gated processes.---}We start by modeling a gated process consisting of two independent components. First, an underlying process $X_{n_0}(t)$, initially at $n_0$, modeled as a continuous-time Markov process. Second, a gate modeled by a two-state continuous-time Markov process, that intermittently switches between an `open' active ($A$) state and a `closed' inactive ($I$) state.  This gate accounts for the additional constraint that needs to be satisfied for the task of interest to be completed. The gate switches from state $A$ to $I$ at rate $\alpha$, and from $I$ to $A$ at $\beta$. For 
$\sigma_0, \sigma \in \{A,I\}$, we define  $p_{t}(\sigma|\sigma_0)$ to be the probability that the gate is in state $\sigma$ at time $t$, given that it was in state $\sigma_0$ initially (see SI for an explicit formula \cite{SI}). Also, let $\pi_A=\beta / \lambda$ and $\pi_I=\alpha / \lambda$ denote the equilibrium occupancy probabilities of states $A$ and $I$ respectively, where $\lambda=\alpha+\beta$ is the relaxation rate to equilibrium. 

The central quantity of interest in our Letter is the first-passage time $T_f(m|n_0)$, which is the time taken for $X_{n_0}(t)$ to reach state $m$ for the first time, and we denote its probability density by $F_t(m|n_0)$. In many scenarios the first-passage time is not directly measurable, and instead we can only measure  the detection time $T_d(n_0,\sigma_0)$, of a reaction or threshold crossing event. We denote by $D_t(n_0,\sigma_0)$ the probability density of $T_d(n_0,\sigma_0)$, which is  the first time the underlying process is \emph{detected} in some target-set $\mathcal{Q}$, given that the initial state of the composite process (underlying + gate) is initially at $\{n_0,\sigma_0\}$. 

In this work, we focus on two widely applicable settings: (i) the detection of threshold crossing events of a 1-dimensional intermittent time-series with nearest-neighbor transitions, where $\mathcal{Q}$ denotes all states above a certain threshold $m$ and $T_d(n_0,\sigma_0)$ is the first time when $X_{n_0}(t)\geq m$ while the detector is active (A), and (ii) gated reactions or target search on an arbitrary network in discrete space or in arbitrary dimension in continuous space. Here, $\mathcal{Q}$ is typically a single target state/point $m$, and $T_d(n_0,\sigma_0)$ denotes the first time the underlying process $X_{n_0}(t)$ is at $m$, while the gate is open (A).

\emph{First-passage times from gated observations.---} We begin our analysis by noting that  for $n_0 \not\in \mathcal{Q}$ we have
\begin{align}
 \nonumber   D_t(n_0,&\sigma_0) =F_t(m|n_0)p_t(A|\sigma_0) ~+ \\ & \int_{0}^{t}F_{t'}(m|n_0)p_{t'}(I|\sigma_0) D_{t-t'}(m,I)~dt',
 \label{eq1}
\end{align}
where the probability for a detection event occurring at time $t$ has two contributions: (i) the detection time coinciding with the first-passage time, and (ii) the gate being closed during the first-passage event (I), and detection happening strictly after this moment in time. The Laplace transform of Eq.~\eqref{eq1}, can be expressed in compact form as \cite{SI}
\begin{align}
    \label{eq2}\widetilde{D}_s(n_0,& \sigma_0) = \left[\pi_A +\pi_I \widetilde{D}_{s}(m,I)\right] \widetilde{F}_s(m|n_0)\\
    & +\mathbb{1}(\sigma_0) (1-\pi_{\sigma_0}) \left[ 1- \widetilde{D}_{s}(m,I)\right]\widetilde{F}_{s+\lambda}(m|n_0), \nonumber
\end{align}
where $\lambda = \alpha + \beta$, and $\mathbb{1}(\sigma_0)$ takes values $+1$ or $-1$ when $\sigma_0 = A$ or $I$, respectively. By explicitly writing down the equations for $\sigma_0 = A$ and $I$, and further eliminating $\widetilde{F}_{s+\lambda}(m|n_0)$ from the equations, we arrive at \cite{SI}
\begin{equation}
    \widetilde{F}_s(m|n_0) = \frac{\pi_A\widetilde{D}_s(n_0,A)+\pi_I\widetilde{D}_s(n_0,I)}{\pi_A+\pi_I\widetilde{D}_s(m,I)},
    \label{centres}
\end{equation}
which is our first result. Equation \eqref{centres} asserts that the first-passage density can be obtained exactly in terms of detection time densities and the gating rates. In \cite{SI}, we further show that Eq.~\eqref{centres} holds even when the underlying process in not Markovian, and instead is a renewal process. However, inference of the first-passage time density $F_t(m|n_0)$, requires  the detection statistics with initial conditions $\{n_0,A\}$, $\{n_0,I\}$, $\{m,I\}$ \emph{and} the equilibrium probabilities $\pi_A$ and $\pi_I$. Such information may not be accessible in experimentally realizable scenarios where, e.g., it may not be possible to initialize a gated molecule in a specific internal state $\sigma_0 = A$ or $I$, and the values of $\pi_A$ and $\pi_I$ may also be unknown. 

In such situations, the most practically realizable initial condition is the equilibrium $\sigma_0 \equiv E$, where the gate is in the active state $A$ with probability $\pi_A$, and in the inactive state $I$ with probability $\pi_I$. Note that this initial condition is naturally achieved if the system is simply allowed to equilibriate. Interestingly, the detection time density starting from the initial condition $(n_0,E)$ is given by $D_t(n_0,E) = \pi_A\cdot D_t(n_0,A) + \pi_I\cdot D_t(n_0,I)$, whose Laplace transform is the numerator standing on the right-hand side of Eq.~\eqref{centres}. Further noting that the Laplace transform of $D_t(m,E) = \pi_A\cdot \delta(t) + \pi_I\cdot D_t(m,I)$ gives the denominator on the right-hand side, we obtain an elegant reinterpretation of Eq.~\eqref{centres}:
\begin{equation}
\widetilde{F}_s(m|n_0) = \frac{\widetilde{D}_s(n_0,E)}{\widetilde{D}_s(m,E)}.
\label{centres_rxn}
\end{equation}
Strikingly, Eq.~\eqref{centres_rxn} asserts that the first-passage time density can be inferred from the detection statistics, even without the explicit knowledge of $\pi_A$ and $\pi_I$, or control over the initial state of the gate.

\begin{figure}[t]
    \centering    \includegraphics[width=0.95\columnwidth,height=5.3cm]{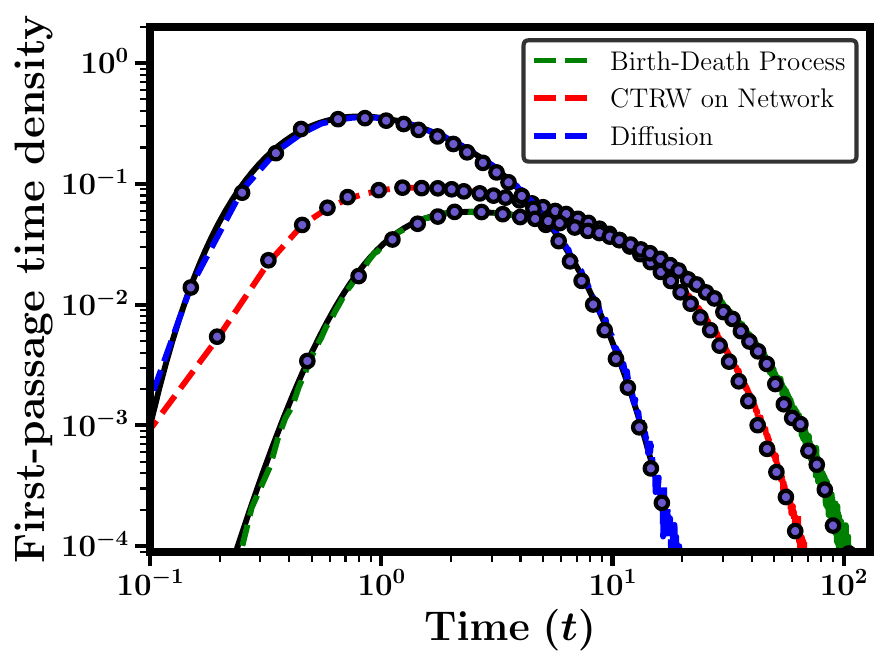}
    \caption{Inference of first-passage time distributions from gated observations. We consider the three different settings mentioned in the text and legend. Solid and dashed lines denote ungated first-passage time distributions obtained from theory and simulations, respectively. Circles are inferred using Eq.~\eqref{centres_rxn} from  histograms of simulated gated detection times \cite{SI}.}
    \label{fig:fptd_infer}
\end{figure}

The usefulness and validity of Eq.~\eqref{centres_rxn} is demonstrated in Fig.~\ref{fig:fptd_infer}, with the help of three case studies of wide interest and applicability. First, a Markovian birth-death process which has been extensively used to model threshold activated reactions \cite{bdapprox,greben1,grebenkov_reversible_2022,grebenkov2022first} and the dynamics of chemical reactions on catalysts \cite{chaudhury2020theoretical,punia_understanding_2021}. Second, the paradigmatic continuous-space diffusion in a 1D confinement. Third, a gated chemical reaction/target search modeled by a non-Markovian continuous-time random walk (CTRW) \cite{montroll1965random,klafter2011first} on a network \cite{masuda2017random}, which is e.g., used to model the motion of reactants, cells, or organisms in complex environments \cite{scher_unified_2021,masuda2017random,metzler2000random,kang2011spatial,hofling2013anomalous,grebenkov2018heterogeneous,berkowitz1997anomalous,berkowitz1998theory}. In all of these settings, we show that the first-passage time distributions inferred from Eq.~\eqref{centres_rxn} using a procedure described in \cite{SI} (circles) are  in excellent agreement with the true first-passage time distributions. We stress  that this inference was performed solely using detection time histograms obtained from gated simulations, without assuming knowledge of their analytical expressions or model specific details (e.g. the network structure and the waiting time distribution in the CTRW example). However, when analytical expressions are available, like in the case of the birth-death process \cite{kumar_first_2021}, one can directly perform the inference through Laplace inversion of Eq.~\eqref{centres_rxn} \cite{SI}.

Before moving forward, we note that Eq.~\eqref{centres_rxn} is reminiscent of the seminal renewal formula $\widetilde{F}_s(m|n_0) = \frac{\widetilde{P}_s(m|n_0)}{\widetilde{P}_s(m|m)}$  which relates, in Laplace space, the first-passage time density and the probability density $P_t(n_i|n_j)$ of finding the underlying process in state $n_i$ at time $t$, given its initial state $n_j$ \cite{redner2001}. Clearly, the right-hand side of this formula and that of Eq.~\eqref{centres_rxn} are equal. In fact, we can obtain an even more general relation -- considering two different initial states $n_0$ and $n'_0$, and after some algebra, we uncover the fundamental relation \cite{SI} 
\begin{equation}
    \frac{\widetilde{D}_s(n_0,E)}{\widetilde{D}_s(n'_0,E)} = \frac{\widetilde{P}_s(m|n_0)}{\widetilde{P}_s(m|n'_0)},  \label{detect_prop}
\end{equation}
asserting that the ratio of the detection time densities (in Laplace space), starting from any two initial states $n_0$ and $n'_0$, is independent of the gating rates $\alpha$ and $\beta$. Note that this is true despite the fact that the detection time densities themselves depend on the gating rates. We remark that Eq.~\eqref{detect_prop} holds in both settings: when $D_t(n_0,E)$ corresponds to gated target search and to the detection of threshold crossing events under intermittent sensing.  

\emph{Inferring the mean first-passage time.---}The Laplace transform in Eq.~\eqref{centres_rxn} allows us to obtain all moments of the first-passage time in terms of moments of the detection time. Equation~\eqref{centres_rxn} further implies that all cumulants of the first-passage time can be expressed as differences between cumulants of detection times. For example, the mean first-passage time is given by 
\begin{align}
\langle T_f(m|n_0)\rangle = \langle T_d(n_0,E)\rangle - \langle T_d(m,E)\rangle.\label{mean}
\end{align}
While simple, Eq.~\eqref{mean} carries utmost importance in practical scenarios, where reliably estimating the full probability distribution is not a viable option, and only the mean can be accurately measured. Apart from setting an important time-scale for a wide class of chemical reactions in confinement, where the mean reaction time can be used to infer full reaction time statistics \cite{guerin2021universal}, the mean first-passage time can also shed light on fundamental properties of the system at hand \cite{belousov_first-passage_2020,belousov2022statistical}. 

\emph{Inferring the diffusion coefficient.---}We now illustrate how one can utilize our framework to infer physically  meaningful parameters like the diffusion coefficient $\mathcal{D}$. Importantly, we show that this can be done even when the actual motion of the particle cannot be tracked. Imagine a scenario like that depicted in Fig.~\ref{fig:schematic}(b), namely we inject an unobservable particle---whose detection is possible only upon reaction---at a known location $x_0$. Assume that the internal state of the particle is initially equilibrated ($\sigma_0=E$); and further assume that it is freely diffusing inside an effectively one-dimensional box $[0,L]$ with reflecting boundaries and a gated target located at $x_0<m<L$. Utilizing Eq.~\eqref{mean} we find that \cite{SI} 
\begin{equation}
\mathcal{D} = \frac{1}{2}\frac{m^2 - x_0^2}{\langle T_d(x_0,E)\rangle - \langle T_d(m,E)\rangle}.
\label{extractdiffusion}
\end{equation}
Equation~\eqref{extractdiffusion} asserts that the diffusion coefficient can be inferred from the difference in the measurable detection times $\langle T_d(x_0,E)\rangle$ and $\langle T_d(m,E)\rangle$.

\begin{figure} 
    \centering
    \includegraphics[width=0.97\linewidth]{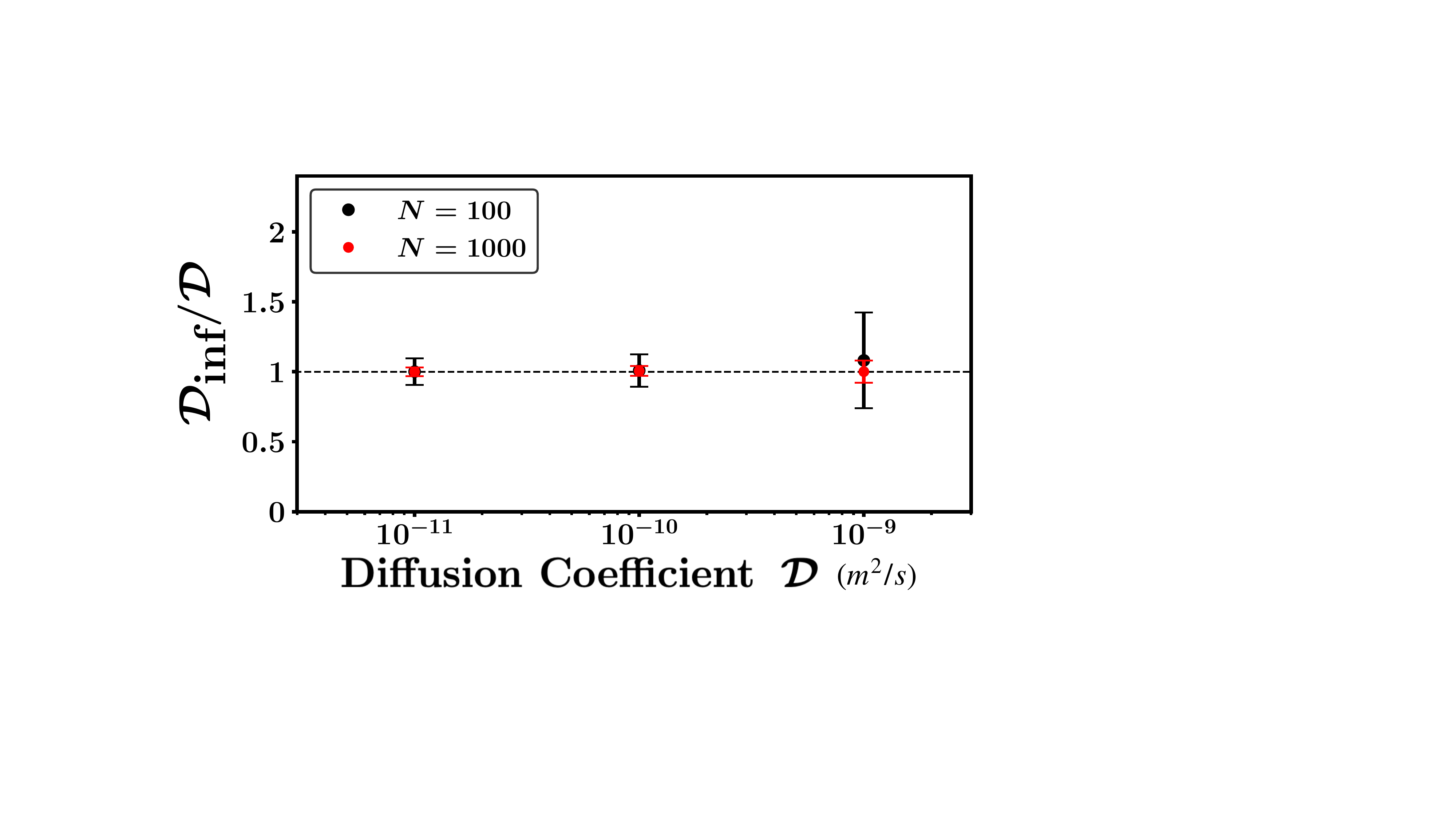}
    \caption{Inference of the diffusion coefficient. Equation \eqref{extractdiffusion} is used to infer the diffusion coefficient of an  unobservable particle that is injected at a known location $x_0=0$ into a box $[0,5 \mu m]$ with reflecting boundaries. The initial internal state is equilibrated $\sigma_0=E$, and a gated point target is located at $m=4 \mu m$, with gating rates $\alpha=\beta=10^2 s^{-1}$. }
    \label{fig:diffusion_inference}
\end{figure}

To corroborate this finding, we simulate the aforementioned scenario and test it for a wide range of possible diffusion coefficients (Fig.~\ref{fig:diffusion_inference}). 
As implied by Eq.~\eqref{mean}, the difference in the detection times is independent of the transition rates, the box size $L$, and the target size (the same equation will hold for threshold crossing). It is thus up to the experimentalist to tune these parameters such that the detection times can be measured with sufficient accuracy. Here we set $\alpha=\beta=10^2 s^{-1}$ and $L=5 \mu m$. For each value of $\mathcal{D}$, the corresponding mean detection times were estimated from averages of $N=10^2$ and $10^3$ simulations, and the diffusion coefficient was inferred via Eq.~\eqref{extractdiffusion}. The errors bars were estimated by repeating this procedure $10^2$ times and noting the standard deviation. In Fig.~\ref{fig:diffusion_inference} we plot the ratio between the inferred values and the actual ones.  We find this estimation procedure robust, even when the number of measurements is relatively small ($N=10^2$). For the parameters used here, the estimation is especially accurate for smaller diffusion coefficients, where mean detection times are longer.

\emph{Inferring the gating rates.---}Equation ~\eqref{centres_rxn} states that the first-passage time density can be inferred from its gated counterparts, even without any prior knowledge of the gating rates $\alpha$ and $\beta$ or control over the initial internal condition. We will now illustrate how the inferred first-passage time distribution can be used together with the observed  detection time distribution to infer the gating rates, thus providing insight into the dynamics of the gating process. 

To proceed, we shift our focus to short-time asymptotics analysis which, despite several recent applications in stochastic thermodynamics \cite{das2022inferring,manikandan2021quantitative,van2022thermodynamic} and chemical kinetics \cite{thorneywork_direct_2020,li2013mechanisms}, has not yet been used to further our knowledge on gated processes. In the short-time limit, the dominant contribution to $D_t(n_0,E)$ comes from trajectories where the detection occurs upon first arrival. This insight translates to the limiting equation $\pi_A = \lim_{t \to 0} D_t(n_0,E)/F_t(m|n_0)$. Similarly, the short-time asymptotics of $D_t(m,E)$ is given by $\pi_I = \beta^{-1} \lim_{t \to 0}  D_t(m,E)$,
owing to the fact that when the underlying process starts on $m$, the dominant contribution to detection comes from events where the gate opens before the particle leaves the target or falls below the threshold. 

These limiting representations of the probabilities $\pi_A$ and $\pi_I$, along with their normalization, allows us to obtain the gating rates as follows
\begin{equation}
 \alpha = \lim_{t \to 0} \frac{D_t(m,E) F_t(m|n_0)}{D_t(n_0,E)},
    \label{alpha}
\end{equation} 
\begin{equation}
     \beta = \lim_{t \to 0} \frac{D_t(m,E) F_t(m|n_0)}{F_t(m|n_0) - D_t(n_0,E)}.  \label{beta}
\end{equation}
\begin{figure} 
    \centering
    \includegraphics[scale=0.19]{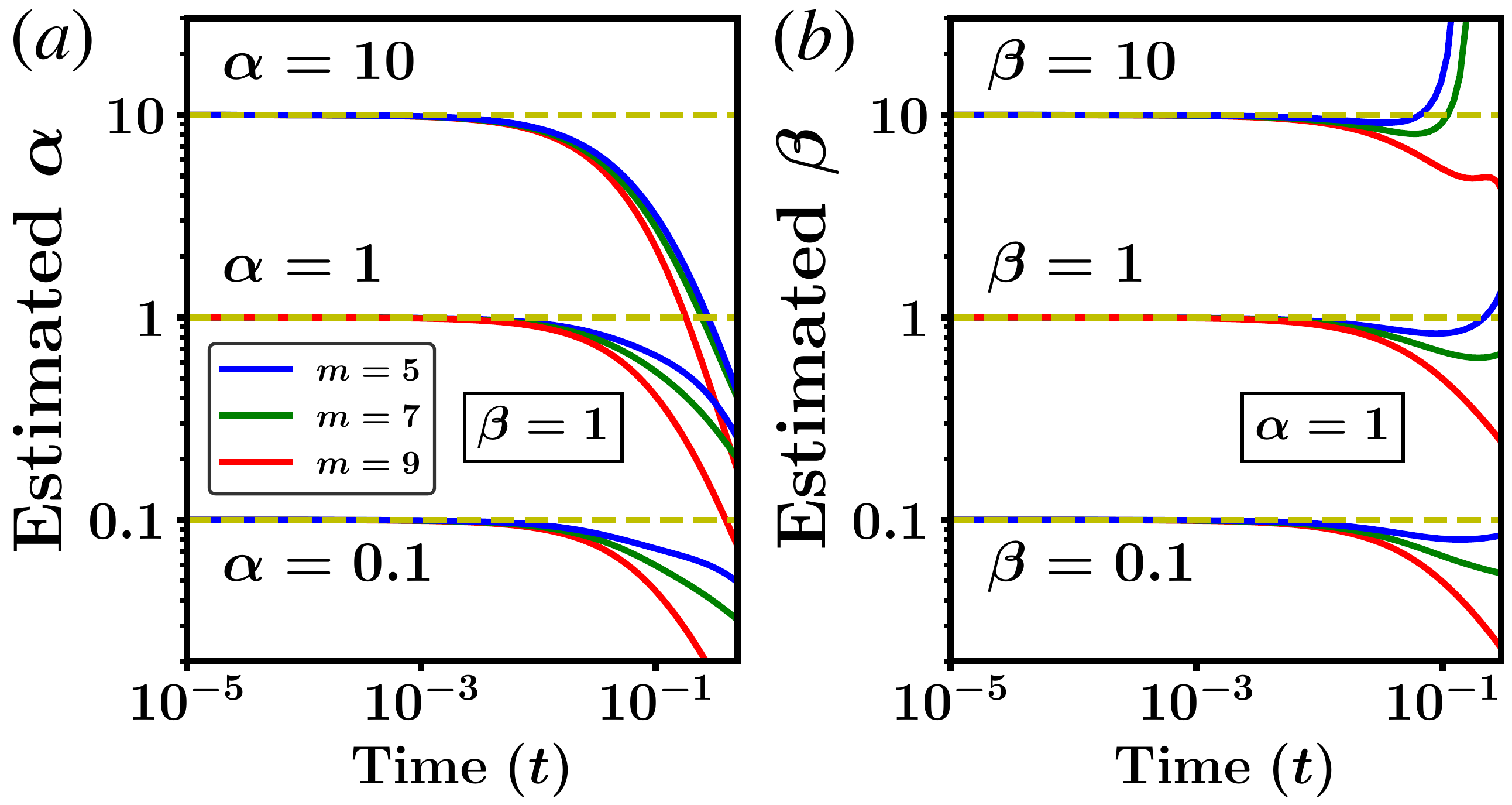}
    \caption{Inference of the gating rates $\alpha$ (panel a) and $\beta$ (panel b) from the short-time asymptotics of Eq.~\eqref{alpha}~and~\eqref{beta} respectively. Results are for  the birth-death model used in Fig.~\ref{fig:fptd_infer}, and various values of $\alpha$ and $\beta$. Details of the model and parameter values are given in \cite{SI}.}
    \label{fig:gate_rate}
\end{figure}

Equations~\eqref{alpha} and~\eqref{beta} are corroborated in Fig.~\ref{fig:gate_rate}, for the birth-death process with parameters described in \cite{SI}. Furthermore, in \cite{SI} we also show that these relations hold even for an arbitrary (non-equilibrium) initial condition of the gate. We then derive simpler inference relations for the gating rates, which are obtained at the cost of perfect control over $\sigma_0$. Finally, we discuss the widely applicable case of simple diffusion and derive inference relations for $\alpha$ and $\beta$, which only differ by a factor of $2$ from Eqs.~\eqref{alpha}~and~\eqref{beta}. 

\emph{Discussion.---}Using the unified framework of gated first-passage processes, we demonstrated how the first-passage time distribution can be inferred from gated measurements, and using these quantities, key features of the process can be extracted. The exact results obtained in this Letter can help inform statistical inference frameworks designed to deal with situations pertaining to imperfect observation conditions, including sparsely sampled time-series or missing data problems. The asymptotic results presented in this Letter moreover provide a systematic approach to the inference of gating rates which, depending on the accessible timescales of the problem, can be improved upon by considering higher-order corrections to the asymptotics.

\emph{Acknowledgments.---} The authors thank Ofek Bonomo-Lauber for carefully reading and commenting on the manuscript, and Ritam Pal for fruitful discussions. A.K. gratefully acknowledges the Prime Minister’s Research Fellowship of the Government of India for financial support. M.S.S. acknowledges the support of a MATRICS Grant from SERB, Government of India. S.R acknowledges support from the Israel Science Foundation (grant No. 394/19). This project has received funding from the European Research Council (ERC) under the European Union’s Horizon 2020 research and innovation program (Grant agreement No. 947731).

\noindent A.K. and Y.S. contributed equally to this work.

%

\clearpage

\onecolumngrid

\hspace{8ex}{\large \textbf{Supplemental Material for ``Inference from gated first-passage times"}}

\vspace{5ex}

This Supplemental Material provides further discussion and derivations which support our findings reported in the Letter, and provides details of the models and simulations used to validate our results. 


\setcounter{page}{1}
\renewcommand{\thepage}{S\arabic{page}}
\setcounter{equation}{0}
\renewcommand{\theequation}{S\arabic{equation}}
\setcounter{figure}{0}
\renewcommand{\thefigure}{S\arabic{figure}}
\setcounter{section}{0}
\renewcommand{\thesection}{S\arabic{section}}
\setcounter{table}{0}
\renewcommand{\thetable}{S\arabic{table}}

\section{S1. Dynamics of the gate: explicit formula for $p_{t}(\sigma|\sigma_0)$} \label{S1}

The gate is modelled as a two-state Markov process, which switches from state $A$ to $I$ at rate $\alpha$, and from $I$ to $A$ at rate $\beta$. For $\sigma_0, \sigma \in \{A,I\}$, we define  $p_{t}(\sigma|\sigma_0)$ to be the probability that the gate is in state $\sigma$ at time $t$, given that it was in state $\sigma_0$ initially, and let $\pi_A=\beta / \lambda$ and $\pi_I=\alpha / \lambda$ denote the equilibrium occupancy probabilities of states $A$ and $I$ respectively, where $\lambda=\alpha+\beta$ is the relaxation rate to equilibrium.
The internal dynamics propagator $p_{t}(A|A)$ is thus governed by the following differential equation

\begin{equation} 
\frac{d p_{t}(A|A)}{d t} = -\alpha p_{t}(A|A) + \beta \left(1-p_{t}(A|A)\right),
\end{equation}

\noindent and from normalization we have $p_{t}(I|A)=1-p_{t}(A|A)$. Similarly,

\begin{equation} 
\frac{d p_{t}(A|I)}{d t} = -\alpha p_{t}(A|I) + \beta \left(1-p_{t}(A|I)\right),
\end{equation}

\noindent and $p_{t}(I|I)=1-p_{t}(A|I)$. The solutions for these differential equations are
\begin{equation} \label{internal_propagator}
  \begin{array}{ll}
   p_t(A \mid I) =\pi_{A}(1-e^{-\lambda t}) ,     \\
   p_t(I \mid I) =\pi_{I}+\pi_{A}e^{-\lambda t}  ,     \\
   p_t(A \mid A) =\pi_{A}+\pi_{I}e^{-\lambda t},     \\
   p_t(I \mid A) =\pi_{I}(1-e^{-\lambda t}).  
  \end{array}
\end{equation}
It is evident that in the long time limit, these probabilities tend to the corresponding equilibrium occupancy probabilities, \emph{i.e,} $\lim_{t\to \infty} p_t(A|A) = \lim_{t\to \infty} p_t(A|I) = \pi_A$ and $\lim_{t\to \infty} p_t(I|A) = \lim_{t\to \infty} p_t(I|I) = \pi_I$.
\section{S2. First-passage times from gated measurements: derivation of Eqs.~(2-4) of the main text} \label{S2}

\begin{figure}
    \centering
    \includegraphics[width = 0.5\columnwidth]{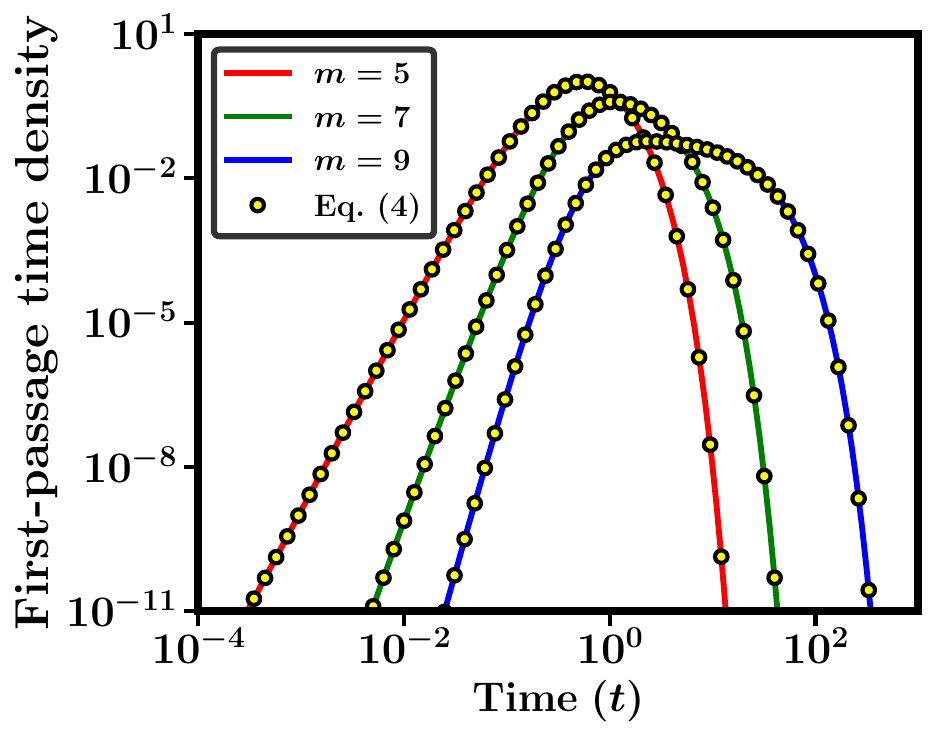}
    \caption{First-passage time distribution for the birth-death process used in the text for three different values of the threshold. Solid lines denote the true first-passage time distribution and symbols (yellow circles) denote the inferred distribution from a numerical Laplace inversion of the ratio of the detection time densities on the right hand side of Eq.~\eqref{SI_s12}. The parameter values chosen for the birth-death process are $N=10$ and $k_+ = k_- = 1$.}
    \label{fig:SI_infer1}
\end{figure}

We recall that Eq.~(1) of the main text states
\begin{equation} 
   D_t(n_0,\sigma_0)=F_t(m|n_0)p_t(A|\sigma_0) + \int_{0}^{t}F_{t'}(m|n_0)p_{t'}(I|\sigma_0) D_{t-t'}(m,I) dt', \label{SI_S4}
\end{equation}
where $n_0$ denotes the initial state of the underlying process, and $m$ denotes the target state (in the context of gated target search), or a threshold (in the context of gated threshold crossing). It is to be noted that Eq.~\eqref{SI_S4} is valid when $n_0\neq m$ in gated target search, and $n_0 < m$ for the threshold crossing problem. Setting $\sigma_0 = A$ we plug in Eq.~\eqref{SI_S4} the relevant expressions from 
Eq. (\ref{internal_propagator}). Then, moving to Laplace space, the Laplace transform of the first term on the right-hand side reads
\begin{equation}
   \mathcal{L} \{F_t(m|n_0)p_t(A|A)\} = \int_0^{\infty} F_t(m|n_0) \left( \pi_{A}+\pi_{I}e^{-\lambda t} \right) e^{-s t}  dt = \pi_{A}\widetilde{F}_{s}(m|n_0) +  \pi_{I}\widetilde{F}_{s+\lambda}(m|n_0),
\end{equation}
and the second term, which is a convolution (denoted by $*$ for brevity), reads
\begin{align}
 \mathcal{L} \{ \left[F_{t}(m|n_0) \pi_{I}(1-e^{-\lambda t})\right] *  D_{t}(m,I) \} & = \mathcal{L} \{ F_{t}(m|n_0) \pi_{I}(1-e^{-\lambda t})\}  \mathcal{L} \{ D_{t}(m,I) \}    \\
&   = \pi_{I}\left(\widetilde{F}_{s}(m|n_0) - \widetilde{F}_{s+\lambda}(m|n_0) \right) \widetilde{D}_{s}(m,I). \nonumber
\end{align}
Put together, we have
\begin{align}
    \widetilde{D}_s(n_0,A) = \pi_{A}\widetilde{F}_{s}(m|n_0) +  \pi_{I}\widetilde{F}_{s+\lambda}(m|n_0) + \pi_{I}\left(\widetilde{F}_{s}(m|n_0) - \widetilde{F}_{s+\lambda}(m|n_0) \right) \widetilde{D}_{s}(m,I),
\end{align}
or
\begin{align}
    \widetilde{D}_s(n_0,A) = \Big(\pi_{A}+\pi_I \widetilde{D}_s(m,I)\Big)\widetilde{F}_{s}(m|n_0) +  \pi_{I}\Big(1- \widetilde{D}_s(m,I)\Big)\widetilde{F}_{s+\lambda}(m|n_0).\label{SI_s8}
\end{align}
Similarly, the corresponding equation for  $\sigma_0=I$ reads
\begin{align}
    \widetilde{D}_s(n_0,I) = \Big(\pi_{A}+\pi_I \widetilde{D}_s(m,I)\Big)\widetilde{F}_{s}(m|n_0) -  \pi_{A}\Big(1- \widetilde{D}_s(m,I)\Big)\widetilde{F}_{s+\lambda}(m|n_0).\label{SI_s9}
\end{align}
It is easy to see that Eqs.~\eqref{SI_s8}~and~\eqref{SI_s9} can be written in compact form as
\begin{align}
    \widetilde{D}_s(n_0,& \sigma_0) = \left[\pi_A +\pi_I \widetilde{D}_{s}(m,I)\right] \widetilde{F}_s(m|n_0)
    +\mathbb{1}(\sigma_0) (1-\pi_{\sigma_0}) \left[ 1- \widetilde{D}_{s}(m,I)\right]\widetilde{F}_{s+\lambda}(m|n_0),
\end{align}
where $\mathbb{1}(\sigma_0)$ takes values $+1$ or $-1$ when $\sigma_0 = A$ or $I$, respectively, yielding Eq.~(2) of the main text. Furthermore, the term containing $\widetilde{F}_{s+\lambda}(m|n_0)$  can be eliminated from Eqs.~\eqref{SI_s8}~and~\eqref{SI_s9} by multiplying them by $\pi_A$ and $\pi_I$ respectively, and adding the two resulting equations. This allows us to write 
\begin{align}
        \widetilde{F}_s(m|n_0) = \frac{\pi_A\widetilde{D}_s(n_0,A)+\pi_I\widetilde{D}_s(n_0,I)}{\pi_A+\pi_I\widetilde{D}_s(m,I)},
        \label{SI_s11}
\end{align}
which identifies with Eq.~(3) in the main text and completes our proof that the first-passage time density $F_t(m|n_0)$ can be expressed in terms of the gated detection time densities, and the gate steady-state probabilities $\pi_A$ and $\pi_I$. We note that the detection time density of an underlying process initially in state $n_0$ and with a gate initially equilibrated, $D_t(n_0, E)$, is nothing but a weighted average of detection time densities starting from an active and inactive gate respectively: $D_t(n_0, E) = \pi_A D_t(n_0,A) +\pi_I D_t(n_0,I)$. Similarly, $D_t(m, E) = \pi_A \delta(t) +\pi_I D_s(m,I)$. This allows us to write Eq.~\eqref{SI_s11} as
\begin{align}
        \widetilde{F}_s(m|n_0) = \frac{\widetilde{D}_s(n_0,E)}{\widetilde{D}_s(m,E)}
        \label{SI_s12}
\end{align}
which can be identified with Eq.~(4) from the main text. In Fig.~\ref{fig:SI_infer1}, we consider the example of the birth-death process discussed in the main text. We show that the first-passage time distribution inferred from a numerical Laplace inversion of the ratio of the detection time densities on the right hand side of Eq.~\eqref{SI_s12} (yellow circles) agrees with the true first-passage time distribution.

\subsection{A. Extension to renewal processes}

In the above derivation we assumed that the underlying spatial process is a continuous-time Markov process. However, this modelling assumption can be relaxed, and our results can be shown to be valid even when the underlying process is modelled as a renewal process. 

To see this, we use as a starting point the following renewal structure \citep{1}
\begin{equation}  \label{renewal_PRL_SI}
T_d(n_0,\sigma_0)= T_f(m| n_0) +
    \begin{cases}
      0, &\text{if~~} \sigma_{T_f(m|x_0)}= A\\
       T_d(m,I), &\text{otherwise},
    \end{cases}
\end{equation}
where $T_d(n_0,\sigma_0)$, $T_d(m,I)$, and $T_f(m|n_0)$ denote random variables whose densities have been denoted by $D_t(n_0,\sigma_0)$, $D_t(m,I)$, and $F_t(m|n_0)$ in the main text. By $\sigma_{T_f(m|x_0)} = A$ in the above equation, we mean that the random variable denoting the state of the gate takes value $A$ at time $T_f(m|n_0)$, given that its state initially was $\sigma_0$. Equation~\eqref{renewal_PRL_SI} is valid even when the underlying spatial process is an arbitrary renewal process \citep{1} and thus allows us to relax the Markovian assumption.

Taking a Laplace transform of Eq.~\eqref{renewal_PRL_SI}, and rearranging, we get
\begin{align}
    \widetilde{D}_s(n_0, \sigma_0) = \left[\pi_A +\pi_I \widetilde{D}_{s}(m,I)\right] \widetilde{F}_s(m|n_0)
    +\mathbb{1}(\sigma_0) (1-\pi_{\sigma_0}) \left[ 1- \widetilde{D}_{s}(m,I)\right]\widetilde{F}_{s+\lambda}(m|n_0), 
\end{align}
which is the same as Eq.~(2) in the main text. Hence, Eq.~(3) in the main text follows for renewal processes as well, using the exact same procedure.

\section{S3. Inference of first-passage times from data based on Eq. (4)}

\begin{figure*}
    \centering
    \includegraphics[width=\columnwidth]{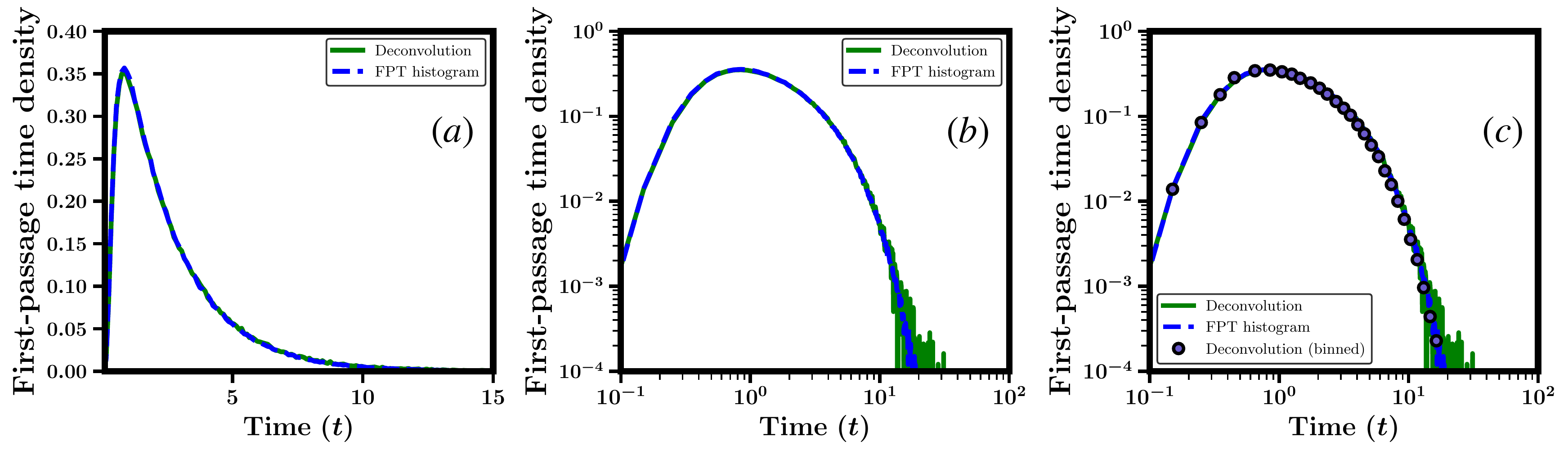}
    \caption{Illustration of the deconvolution method to infer the first-passage statistics from gated detection time data, using the example of a diffusing particle confined in a $1$D box. (a) The deconvoluted first-passage time density overlaps almost perfectly with the FPT histogram computed from ungated simulations. (b) A log-log plot shows that there are numerical errors when estimating very small   probabilities (the noisy green tail). However, by taking logarithmic bins and assigning the mean value of the densities within each bin to be the density for that bin, we show in panel (c) that the full circles  (depicting the mean values of the values in each bin) match the FPT histogram well.}
    \label{fig:SI_infer2}
\end{figure*}

Equation~(4) in the main text (Eq.~\eqref{SI_s12} in the SI) allows us to represent the first-passage time distribution purely in terms of the observable detection time distributions.  In this section, we show that this equation can be used to infer the first-passage time distribution directly from detection times data. This is of utmost importance, since in many practically relevant scenarios, analytical expressions for the detection time distributions are not known as we might not know the laws of motion or specific parameter values of the underlying process. First, we note that in the time domain Eq.~\eqref{SI_s12} can be written as a convolution
\begin{equation}
    D_t(n_0,E) = \int_0^t F_{t'}(m|n_0) D_{t-t'}(m,E)~dt'.
    \label{conv}
\end{equation}
This suggests that the problem of inferring the first-passage time distribution from gated detection times can be viewed as a deconvolution problem. In practice, when detection time data are obtained from simulations/experiments, we discretize the detection time distributions by binning the data in histograms. Thus, we discretize Eq.~\eqref{conv}, and recast it as a matrix equation 
\begin{equation}
 \vec{D}_t(n_0,E) = \mathbb{D}_t(m,E) \vec{F}_{t}(m|n_0),
\end{equation}
where $\mathbb{D}_t(m,E)$  is a matrix, which is interpretted as an operator that acts on the vector $\vec{F}_t(m|n_0)$ by performing a convolution and giving the vector $\vec{D}_t(n_0,E)$ as output. More specifically, we have
\begin{equation}
\underbrace{\begin{bmatrix} D_0(n_0,E) \\ D_{\Delta}(n_0,E) \\ D_{2\Delta}(n_0,E) \\ D_{3\Delta}(n_0,E) \\ \vdots \\ D_{\mathcal{N}\Delta}(n_0,E) \end{bmatrix}}_{\text{$\vec{D}_t(n_0,E)$}} = \underbrace{\Delta \begin{bmatrix} D_0(m,E) & 0 & 0 & 0 & \dots & 0 \\ D_{\Delta}(m,E) & D_0(m,E) & 0 & 0 & \dots & 0 \\ D_{2\Delta}(m,E) & D_{\Delta}(m,E) & D_0(m,E) & 0 & \dots & 0 \\ D_{3\Delta}(m,E) & D_{2\Delta}(m,E) & D_{\Delta}(m,E) & D_{0}(m,E) \\ \vdots & \vdots & \vdots & & ~\ddots  \\ D_{\mathcal{N}\Delta}(m,E) & D_{(\mathcal{N}-1)\Delta}(m,E) & D_{(\mathcal{N}-2)\Delta}(m,E) & \dots & \dots & D_0(m,E) \end{bmatrix}}_{\text{$\mathbb{D}_t(m,E)$}}  \underbrace{\begin{bmatrix} F_0(m|n_0) \\ F_{\Delta}(m|n_0) \\ F_{2\Delta}(m|n_0) \\ F_{3\Delta}(m|n_0) \\ \vdots \\ F_{\mathcal{N}\Delta}(m|n_0) \end{bmatrix}}_{\text{$\vec{F}_t(m|n_0)$}},
\end{equation} 
where $\Delta$ is the histograms' bin width and $\mathcal{N}+1$ as the total number of bins. Evidently, $\mathbb{D}_t(m,E)$ is a lower-triangular, Toeplitz, matrix whose entries are densities obtained from the histogram of $D_t(m,E)$, scaled by a factor of $\Delta$. Thus, we can write
\begin{equation}
  \vec{F}_{t}(m|n_0)   = \mathbb{D}_t(m,E)^{-1} \vec{D}_t(n_0,E),
\end{equation}
implying that the problem of inferring the first-passage time distribution reduces to a problem of inverting a lower diagonal Toeplitz matrix with a non-zero determinant. In the main text, we demonstrated the validity and robustness of this approach using three distinct examples -- a birth-death process, CTRW on network, and 1D confined diffusion, where we performed inference from detection time histograms generated from $10^6$ detection events. Further expanding on the diffusion example, in Fig.~\ref{fig:SI_infer2}, we demonstrate the deconvolution method to infer the first-passage statistics from gated detection time data. In panel (a), we clearly note that the deconvoluted first-passage time density shows an excellent agreement with the first-passage time histogram computed from ungated simulations. A corresponding log-log plot in panel (b) reveals some numerical errors in the tail of the inferred first-passage time distribution. In panel (c), we plot the inferred density after a logarithmic binning and we can see that the full circles (depicting the mean values of the values in each bin) match the true first-passage time histogram very well. The logarithmic binning helps in reducing the numerical errors from the tails, however it does not completely remove them. An alternate method to perform this inference could involve parametric inference which uses domain-specific knowledge (e.g., exponential tail of first-passage time distributions in confined systems) to improve upon this inference method.\\
\\

\section{S4. Models used for simulations presented in main text} \label{S4}

\subsection{A. The birth-death process} 

In this section of the Supplemental Material, we describe the birth-death process (BDP) used in the main text of our paper to produce Figs.~2 and 4. 

We considered a BDP on a state space $\mathcal{S} \in \{0,1,2,\cdots,N\}$, with transition rates 
\begin{align}
    \label{transitions1}W_{+}(j)&= k_{+} (N-j),\\
    W_{-}(j)&= k_{-}j,
    \label{transitions2}
\end{align} which govern the rate of transitioning from state $j$ to states $j+1$ and $j-1$ respectively, with $j \in \{0,1,2,\cdots,N\}$.  Clearly, states $0$ and $N$ act as reflecting boundaries, as $W_{-}(0)=W_{+}(N)=0$.

Let us define $F_{t}(m|n_0)$ to be the probability distribution of the time the BDP reaches state $m$ for the first time, starting from state $n_0$. $F_{t}(m|n_0)$ is called the first-passage time density, and it is known that $F_{t}(m|n_0)$ obeys a Phase-type distribution \citep{2a,2b}, whose Laplace transform is easy to compute.

More explicitly, taking the example of $n_0=0$, we note that $F_t(m|0)$ can be simply expressed as 
\begin{equation} 
F_t(m|0) = k_+ (N-m+1) \bigl[\exp(\mathbb{W}^{(m)} t)\bigr]_{m,1} \,.
\end{equation}
where $\mathbb{W}^{(m)}$ is the $m\times m$ matrix obtained by retaining only the
first $m$ columns and $m$ rows of the $N+1 \times N+1$ transition matrix $\mathbb{W}$ containing all zeros, except where \begin{equation*}
\mathbb{W}_{i+1,i} = k_+ (N+1-i)~~~ \text{and}~~~ \mathbb{W}_{i,i+1} = k_- i  ,
\end{equation*}
for $i \in \{1,2,\dots,N \} $, and $\mathbb{W}_{i,i}$ are chosen so that the columns of $\mathbb{W}$ add up to zero.

Alternatively, one can also use the renewal method to determine the Laplace transformed first-passage distribution as 
\begin{equation}
    \widetilde{F}_{s}(m|n_0) = \frac{\widetilde{P}_s(m|n_0)}{\widetilde{P}_s(m|m)}\label{renew},
\end{equation}
where $\widetilde{P}_s(i|j)$ is the Laplace transform of the propagator $P_t(i|j)$, which denotes the probability of finding the BDP in state $i \in \mathcal{S}$ at time $t$, given that the system started from state $j\in\mathcal{S}$ initially. The propagator $P_t(i|j)$ can be obtained for any BDP (see Eq.~(2.4) in Ref.\citep{3}).

The parameters chosen for the production of Fig.~2 in the main text are $N=10$, $m=9$, $k_+ = 0.5$, $k_- = 1$, $\alpha = 2$, and $\beta = 0.5$. For Fig.~4, the parameters are $N=10$, and $k_+ = k_- = 1$, for various values of $m$ and gating rates $\alpha$ and $\beta$.

\subsection{B. Diffusing particle in a closed interval with one reflective boundary and one absorbing boundary}

\noindent We start by computing the Green’s function $C(x,t \mid x_0)$, namely, the conditional probability density function to find the particle at position $x$ at time $t$, given that the initial position is $x_0$, and given a reflective boundary at $0$ and an absorbing boundary at $m$. The initial condition is $C(x, t=0 | x_{0})=\delta\left(x-x_{0}\right)$ by definition, and the boundary conditions we require are a Neumann boundary condition $\frac{\partial C(x, t | x_{0})}{\partial x} \big|_{x=0} = 0$ and a Dirichlet boundary condition $C(x = m, t  | x_{0} ) = 0$. 

Laplace transforming the diffusion equation, we obtain 
\begin{equation} \label{B:1}
   s \tilde{C}(x, s | x_{0}) = \mathcal{D}\frac{d^2 \tilde{C}(x, s | x_{0})}{dx^2}, 
\end{equation}
which is a second-order, linear, homogeneous differential equation.  It has a general solution
\begin{equation} \label{B:2}  
\tilde{C}(x, s | x_{0})=
\begin{cases}
  \tilde{C}_{-}(x, s | x_{0}) = c_{1}(s) e^{x\sqrt{\frac{s}{\mathcal{D}}}} + c_{2}(s) e^{-x\sqrt{\frac{s}{\mathcal{D}}}}, \hspace{5ex} x<x_0    
      \\
  \tilde{C}_{+}(x, s | x_{0}) = c_{3}(s) e^{x\sqrt{\frac{s}{\mathcal{D}}}} + c_{4}(s) e^{-x\sqrt{\frac{s}{\mathcal{D}}}}. \hspace{5ex} x>x_0         
\end{cases}
\end{equation}
Similarly, Laplace transforming the boundary conditions we obtain 
\begin{equation} \label{B:3}
   \frac{d \tilde{C}(x, s | x_{0})}{d x} \big|_{x=0} = 0,
\end{equation}
and
\begin{equation}  \label{B:4}
  \tilde{C}(m, s | x_{0})= 0.
\end{equation}
Finally, the initial condition is translated to two matching conditions at the initial position of the particle, one for the continuity of the Laplace transform of the probability density 
\begin{equation}  \label{B:5}
     \tilde{C}_{+}(x_{0}, s | x_{0}) = \tilde{C}_{-}(x_{0}, s | x_{0}),  
\end{equation}
and one for the Laplace transform of the fluxes 
\begin{equation}  \label{B:6}
     -1=\mathcal{D}\Big[\frac{d \tilde{C}_{+}(x, s | x_{0})}{d x}\big|_{x=x_{0}}  - \frac{d \tilde{C}_{-}(x, s | x_{0})}{d x}\big|_{x=x_{0}}\Big], 
\end{equation}
which is obtained by integrating both sides of the Laplace transformed diffusion equation (Eq.~(\ref{B:1})) over an infinitesimally small interval surrounding the initial position. Note that the $1$ on the left-hand side comes from the Laplace transform of the delta function initial condition. 

Imposing the above conditions produces a system of four equations with four unknowns, from which we can obtain $c_{i}(s)$ ($1 \leq i \leq 4$)

\begin{equation} \label{B:7}
\begin{cases}
   c_{1}(s) = \frac{\text{sech}\left(m \sqrt{\frac{s}{\mathcal{D}}}\right) \sinh \left(\sqrt{\frac{s}{\mathcal{D}}}
   \left(m-x_0\right)\right)}{2 \sqrt{\mathcal{D} s}},
\\
    c_{2}(s) = \frac{\text{sech}\left(m \sqrt{\frac{s}{\mathcal{D}}}\right) \sinh \left(\sqrt{\frac{s}{\mathcal{D}}}
   \left(m-x_0\right)\right)}{2 \sqrt{\mathcal{D} s}}  ,
\\
    c_{3}(s) = \frac{\left(\tanh \left(m \sqrt{\frac{s}{\mathcal{D}}}\right)-1\right) \cosh \left(x_0
   \sqrt{\frac{s}{\mathcal{D}}}\right)}{2 \sqrt{\mathcal{D} s}}  ,
\\
    c_{4}(s) = \frac{e^{m \sqrt{\frac{s}{\mathcal{D}}}} \text{sech}\left(m \sqrt{\frac{s}{\mathcal{D}}}\right) \cosh
   \left(x_0 \sqrt{\frac{s}{\mathcal{D}}}\right)}{2 \sqrt{\mathcal{D} s}}. 
\end{cases}
\end{equation}

The Laplace transform of the first-passage probability density function is given by
\begin{equation}  \label{B:8}
   \tilde{T}_f(m, s) = -\left.\mathcal{D} \frac{d \tilde{C}_{+}(x, s | x_{0})}{d x}\right|_{x=m} 
= \operatorname{sech}\left[m \sqrt{\frac{s}{\mathcal{D}}}\right]\operatorname{cosh} \left[x_0 \sqrt{\frac{s}{\mathcal{D}}}\right] 
\end{equation}

The Laplace transform is a moment generating function, and so the mean first-passage time is given by
\begin{equation}   \label{B:9}
   \braket{T_f}= -\frac{d \tilde{T}_f(m, s)}{d s}\bigg|_{s=0} =\frac{m^2 - x_0^2}{2 \mathcal{D}}.
\end{equation}

The parameter values chosen for the production of Fig.~2 in the main text are $L=2$, $D=\frac{1}{2}$, $x_0=0.2$, $m = 1.6$, and gating rates $\alpha = \beta = 0.5$.

\subsection{C. Continuous-time random walk (non-Markovian) on a network}

A continuous-time random walk (CTRW) on a network is a mathematical framework for modeling the random movement of particles or agents on a network over time. CTRWs on networks have found numerous applications in various fields, including physics, biology, sociology, and computer science. In this model, a random walker, jumps successively from a node of the network to one of its neighbouring nodes after waiting for a random time, drawn from its waiting time distribution. For a comprehensive review, refer to \cite{4}.

\begin{figure}
    \centering
    \includegraphics[width=0.4\columnwidth]{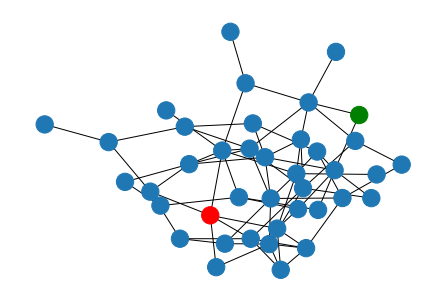}
    \caption{The network used for simulations of gated target search by a CTRW, with the red and green nodes denoting the initial position of the CTRW and the target respectively. For the CTRW on this network, the waiting time distribution was taken to be uniform on the interval $[0,0.2]$, whereas the gating rates were chosen to be $\alpha = \beta = 1$.}
    \label{fig:network}
\end{figure}

For the production of Fig.~2 in the main text, we simulated a CTRW on the network depicted in Fig.~\ref{fig:network} -- an Erdös-Rényi random network with $40$ nodes, where each pair of nodes is connected with probability of $0.1$. For the CTRW on this network, the waiting time distribution was taken to be uniform on the interval $[0,0.2]$, whereas the gating rates were chosen to be $\alpha = \beta = 1$.


\section{S5. The connection between the detection times and propagators: derivation of Eq.~(5) in the main text}

Equation~(4) in the main text (Eq.~\eqref{SI_s12}) asserts that the first-passage time density can be inferred from the detection statistics, even without the explicit knowledge of $\pi_A$ and $\pi_I$, or control over the initial state of the gate. Moreover, Eq.~\eqref{SI_s12} is reminiscent of the seminal renewal formula 
\begin{equation}
\widetilde{F}_s(m|n_0) = \frac{\widetilde{P}_s(m|n_0)}{\widetilde{P}_s(m|m)},    
\end{equation}
which relates, in Laplace space, the first-passage time density and the propagator $P_t(n_i|n_j)$ denoting the probability of finding the underlying process in state $n_i$ at time $t$, given its initial state $n_j$. By equating the two expressions we obtain
\begin{equation}
      \frac{\widetilde{D}_s(n_0,E)}{\widetilde{D}_s(m,E)} =  \frac{\widetilde{P}_s(m|n_0)}{\widetilde{P}_s(m|m)}, \label{SI_s14}
\end{equation}
which remarkably holds for both settings: when $D_t(n_0,E)$ corresponds to gated target search or detection of threshold crossing events under intermittent sensing. 

To explore further, consider the first detection time densities, starting from two arbitrary initial conditions $n_0$ and $n'_0$. From \eqref{SI_s14}, we have the two equations, 
\begin{equation}
      \frac{\widetilde{D}_s(n_0,E)}{\widetilde{D}_s(m,E)} =  \frac{\widetilde{P}_s(m|n_0)}{\widetilde{P}_s(m|m)}~~\text{and}~~\frac{\widetilde{D}_s(n'_0,E)}{\widetilde{D}_s(m,E)} =  \frac{\widetilde{P}_s(m|n'_0)}{\widetilde{P}_s(m|m)}
\end{equation}
whose ratio is simply given by
\begin{equation}
    \frac{\widetilde{D}_s(n_0,E)}{\widetilde{D}_s(n'_0,E)} = \frac{\widetilde{P}_s(m|n_0)}{\widetilde{P}_s(m|n'_0)},
\end{equation}
which is Eq.~(5) in the main text. In fact, this equality can be easily extended to give
\begin{equation}
    \frac{\widetilde{D}_s(n_0,E)}{\widetilde{D}_s(n^{'}_0,E)} = \frac{\widetilde{P}_s(m|n_0)}{\widetilde{P}_s(m|n^{'}_0)} =\frac{\widetilde{F}_s(m|n_0)}{\widetilde{F}_s(m|n'_0)}.
\end{equation}


\section{S6. Inferring the gating rates $\alpha$ and $\beta$}

\subsection{A. Derivation of Eqs.~(8-9)}

In this section, we provide a detailed derivation and discussion of our short-time asymptotic calculations, which allow us to infer $\alpha$ and $\beta$ from the detection time densities in the form of Eqs.~(8)~and~(9) in the main text.

For $\sigma_0 = E$, we have the relation, 
\begin{equation}
    D_t(n_0,E) = \pi_A F_t(m|n_0) + \pi_I \int_{0}^{t} ~F_{t'}(m|n_0) D_{t-t'}(m,I) dt', \label{dtne_SI}
\end{equation}
which can be obtained from Eq.~\eqref{SI_S4}, by noting that $D_t(n_0,E) = \pi_A D_t(n_0,A) + \pi_I D_t(n_0,I)$. Furthermore, we have $D_t(m,E) = \pi_A \delta(t) + \pi_I D_t(m,I)$. Evidently, the short-time behaviour of $D_t(m,E)$ is governed by that of $D_t(m,I)$, which can in turn be expressed as
\begin{equation}
    D_t(m,I) \simeq \beta (1 - \Delta(t))
\end{equation}
where $\Delta(t) \to 0 $ in the $t\to 0$ limit. For example, in the case of gated chemical reactions where $m$ denotes a target state, the short-time expression for $D_t(m,I)$ is
\begin{equation}\label{SI_short_rxn}
    D_t(m,I) \simeq \beta e^{-\beta t} e^{-\gamma_m t} 
\end{equation}
where $\gamma_m$ denotes the rate at which the reactant escapes the target state $m$.  Equation~\eqref{SI_short_rxn} expresses the fact that the dominant contribution to reaction events in the short-time limit comes from the events where the gate opens before the reactant leaves the target state $m$. Using the approximation $e^{-\lambda t} \simeq 1 - \lambda t$, it is easy to see that $\Delta(t) = (\beta + \gamma_m)t + o(t)$, and in the $t\to 0$ limit, we have 
\begin{align}
  \lim_{t\to 0} D_t(m,I) = \beta. \label{dtmi_short_SI}
 \end{align}

Equation~\eqref{dtmi_short_SI} asserts that $D_t(m,I)$ tends to a constant ($\beta$) in the short-time limit. Similarly, the second term of the right-hand side in Eq.~\eqref{dtne_SI} can be safely ignored in the short-time limit, yielding 
\begin{equation}
    D_t(n_0,E) \simeq \pi_A F_t(m|n_0).
\end{equation}
From here, we have
\begin{equation}
    \pi_A = \lim_{t\to 0} \frac{D_t(n_0,E)}{F_t(m|n_0)},\label{pia_SI}
\end{equation}
which appeared in the main text.
In a similar vein, the short-time asymptotics of $D_t(m,E)$ can be expressed as
\begin{equation}
    D_t(m,E) \simeq  \pi_I D_t(m,I).
\end{equation}
Using Eq.~\eqref{dtmi_short_SI}, we arrive at
\begin{equation}
    \pi_I= \lim_{t \to 0} \frac{1}{\beta}D_t(m,E), \label{pii_SI}
\end{equation}
which appeared in the main text. 

We are now in the position to derive relations for the inference of $\alpha$ and $\beta$ from Eqs.~\eqref{pia_SI}~and~\eqref{pii_SI}. To obtain these relations, we will use two identities: (i) $\pi_A + \pi_I = 1$, and (ii) $\pi_A \alpha = \pi_I \beta$. While the first identity is trivially the normalization of occupancy probabilities, the latter can be seen as a statement of detailed balance for the two-state Markov process which models our gate.

Writing Eq.~\eqref{pii_SI} as $\pi_I~\beta = \lim_{t \to 0} D_t(m,E) $, and using $\pi_A \alpha = \pi_I \beta$, we have
\begin{equation}
     \pi_A~\alpha = \lim_{t \to 0} D_t(m,E) ,
\end{equation}
where, $\pi_A$ can be further substituted from Eq.~\eqref{pia_SI}, to obtain  Eq.~(8) in the main text
\begin{equation}
 \alpha = \lim_{t \to 0} \frac{D_t(m,E) F_t(m|n_0)}{D_t(n_0,E)}.
    \label{alpha_SI}
\end{equation}
To obtain the corresponding inference relation for $\beta$, we note that Eq.~\eqref{pii_SI} gives
\begin{equation}
    \beta = \frac{1}{\pi_I} \lim_{t\to 0} D_t(m,E),
\end{equation}
where $\pi_I = \frac{\alpha}{\alpha + \beta}$. This gives us
\begin{equation}
    \beta = \frac{\alpha + \beta}{\alpha} \lim_{t\to 0} D_t(m,E),
\end{equation}
which can be further simplified to obtain
\begin{equation}
    \beta = \bigg(1 + \frac{\beta}{\alpha}\bigg) \lim_{t\to 0} D_t(m,E).
\end{equation}
Substituting for $\alpha$ in the above equation from Eq.~\eqref{alpha_SI}, we get
\begin{equation}
      \beta = \lim_{t \to 0} \frac{D_t(m,E) F_t(m|n_0)}{F_t(m|n_0) - D_t(n_0,E)},  \label{beta_SI}
\end{equation}
which is Eq.~(9) from the main text.


\subsection{B. Inference for  arbitrary initial condition of the gate}

In this subsection, we demonstrate that Eqs.~\eqref{alpha_SI}~and~\eqref{beta_SI} hold more generally, and we can relax the condition that the gate is initially equilibriated. 

Consider a situation where we do not have information about the initial conditions of the gate. Let us denote by $p_A$ and $p_I$ the unknown probabilities for the gate to be initially active $A$ and inactive $I$ respectively, for $p_A \in (0,1)$ and $p_I = 1-p_A$. Note that we do not assume that $p_A = \pi_A$ and $p_I = \pi_I$. Following the intuition developed for Eq.~\eqref{pia_SI} and Eq.~\eqref{pii_SI}, we have the following short-time asymptotic relations,
\begin{equation}
 p_A = \lim_{t\to 0} D_t(n_0)/F_t(m|n_0)
\end{equation}
 and 
\begin{equation}
 p_I = \frac{1}{\beta}\lim_{t\to 0} D_t(m),   
\end{equation}
where we have dropped the notation for gate initialization ($D_t(n_0)$ denotes the detection time density given that the initial state of the underlying process is $n_0$, and the initial preparation of the gate is unknown). The normalization of probabilities dictates that $p_A + p_I =1$. So, adding up the above two equations, we get,
 \begin{equation}
     1 = \lim_{t\to 0} \bigg( \frac{1}{\beta} D_t(m) + \frac{D_t(n_0)}{F_t(m|n_0)} \bigg)
 \end{equation}
yielding:
 \begin{equation}
     \beta = \frac{F_t(m|n_0) ~ D_t(m)}{F_t(m|n_0) - D_t(n_0)}
 \end{equation}
which is the same as Eq.~(9) from the main text. Similarly, Eq.~(8) can be derived in this even more general setting.


\subsection{C. Inference in the presence of control over gate initialization}

In this subsection, we discuss the possibility of inferring $\alpha$ and $\beta$ in the case where we can control the initial state of the gate $\sigma_0$. Using the fact that $p_t(A|A) = \pi_A + \pi_I e^{-\lambda t}$ and $p_t(I|A)= \pi_I - \pi_I e^{-\lambda t}$, we have 
\begin{align}
 \nonumber   D_t(n_0,A) =F_t(m|n_0)(\pi_A + \pi_I e^{-\lambda t}) ~+  \int_{0}^{t}F_{t'}(m|n_0)\left(\pi_I( 1 - e^{-\lambda t'})\right) D_{t-t'}(m,I) dt'.
\label{st1}
\end{align}
In the limit of short-time ($t\to 0$), we can write $e^{- \lambda t} \simeq 1 - \lambda t$, which gives us
\begin{align}
  D_t(n_0,&A)  \simeq F_t(m|n_0)  -\pi_I \lambda t F_t(m|n_0),
\end{align}
where we neglect the second term.
 \begin{figure*}
         \centering
         \includegraphics[width=\textwidth]{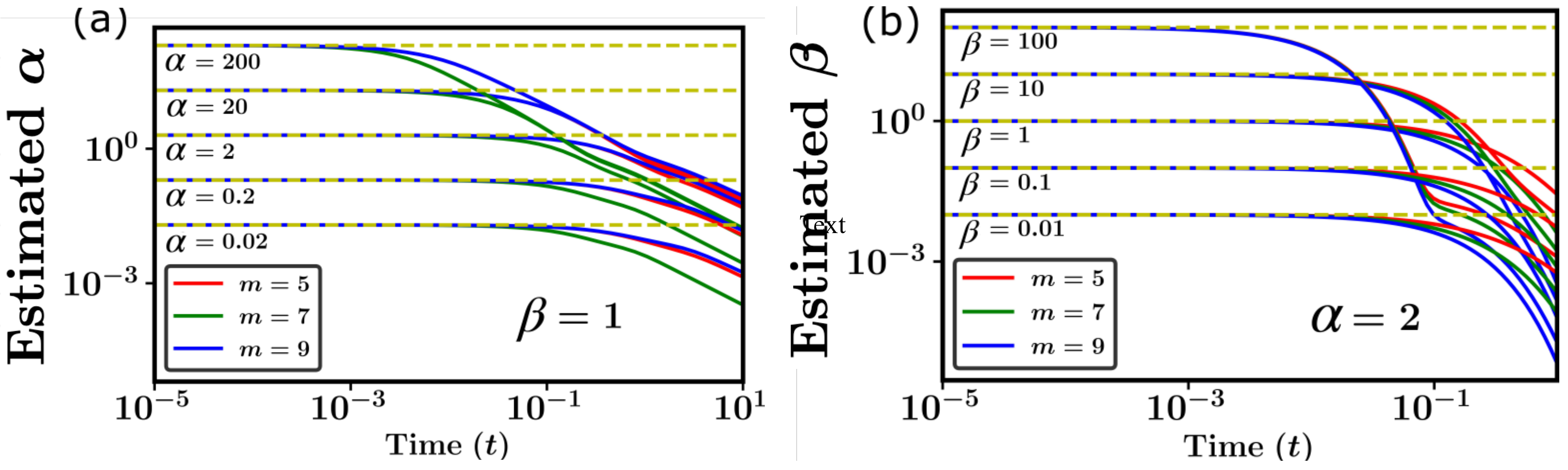}
        \caption{Inference of the gating rates $\alpha$ and $\beta$ from short-time asymptotics for the birth-death process (defined in Sec. S4.A) with $k_+=0.1$, $k_1=1$, and $N=10$. (a) A plot demonstrating that early deviations of the detection time distribution from the first-passage time distribution, as captured by Eq.~\eqref{alpha_control_SI}, can be used to infer the numerical value of $\alpha$. (b) Inference of $\beta$ through the short-time asymptotic behaviour of $D_t(m,I)$ given by Eq.~\eqref{beta_control_SI}. It is clear that the short-time asymptotics of the detection time distributions can be leveraged to infer the gating rates in gated first-passage processes.}
        \label{fig:rate_infer}
\end{figure*}
It is a matter of simple algebra to see that
\begin{align}
     \alpha = \lim_{t \to 0}\frac{F_t(m|n_0) - D_t(n_0,A)}{t \cdot F_t(m|n_0)}.
  \label{alpha_control_SI}
\end{align}
Equation~\eqref{alpha_control_SI} asserts that the early deviations of $D_t(n_0,A)$ from $F_t(m|n_0)$ can be leveraged to obtain the gating rate $\alpha$ corresponding to deactivation of the gate. On the other hand, Eq.~\eqref{dtmi_short_SI} is sufficient to infer $\beta$ as
\begin{align}
\beta = \lim_{t\to 0} D_t(m,I).
\label{beta_control_SI}
\end{align}

In Fig.~\ref{fig:rate_infer}, we corroborate the validity of Eqs.~\eqref{alpha_control_SI}~and~\eqref{beta_control_SI} using the birth-death process described in Sec.~\ref{S4} of this SI (also used in the main text). 
\subsection{D. Inference in the case of diffusion}

An important quantity in the inference of $\alpha$ and $\beta$ from the detection time distributions is the short-time asymptotics of $D_t(m,I)$. This quantity was derived in Eq.~\eqref{dtmi_short_SI} using the fact that, at short-times, the dominant contribution to detection events comes from trajectories where the gate opens before the underlying  process escapes the target-region (i.e., for detection of threshold crossing events, it means that the time-series has not dropped below the threshold, and for gated chemical reactions, it corresponds to the particle not escaping the target site before the opening of the gate).  However, in continuous-space Markov processes this underlying assumption does not hold, since the process can leave and return to the target region multiple times during an infinitesimally small time interval. This means that the second term on the right-hand side of Eq.~\eqref{dtne_SI} can no longer be ignored.

Let us consider the example of a freely diffusing particle in 1-dimension, whose position at time $t$ we denote by $X_{n_0}(t)$ given that it starts from position $n_0$. We define $T_d(n_0,I)$ to be the time taken for the particle to be detected in a location $X_{n_0}(t) \geq m$, given that initially, the gate is in state $\sigma_0 = I$. While Eq.~\eqref{pia_SI} is valid even in this setting, Eq.~\eqref{pii_SI} is not. Thus, in order infer $\alpha$ and $\beta$, our goal is to compute the short-time asymptotics of the probability density $D_t(m,I)$ of the random variable $T_d(m,I)$.

We note that the dominant contribution to $D_t(m,I)$ at short-times comes from the events where detection happens as soon as the gate opens. Note that, in this time, owing to the continuous nature of the process, $X_m(t)$ could have dropped below the threshold $m$ and crossed it subsequently several times. Thus, instead of looking at the first time when it drops below the threshold, the more meaningful question to ask in this setting is: what is the probability that $X_m(t) \geq m$ when the gate opens? Owing to the symmetry of the free diffusion problem, the answer is clearly $\frac{1}{2}$. This observation allows us to write
\begin{equation}
    \lim_{t\to 0} D_t(m,I) = \frac{\beta}{2}, \label{short_threshold}
\end{equation}
and thus
\begin{equation}
    \lim_{t\to 0} D_t(m,E) = \pi_I \lim_{t\to 0} D_t(m,I)  = \pi_I \frac{\beta}{2}.
\end{equation}
Rearranging, we get the analogue of Eq.~\eqref{pii_SI} for continuous processes as
\begin{equation}
    \pi_I = \frac{2}{\beta}\lim_{t\to 0} D_t(m,E), \label{pii_SI_cont}
\end{equation}
where we remark that the right-hand side of Eq.~\eqref{pii_SI_cont} differs from the right-hand side of Eq.~\eqref{pii_SI} only by a factor of 2. This relation can be utilized along with Eq.~\eqref{pia_SI} to obtain
\begin{equation}
      \beta = \lim_{t \to 0} ~\frac{2~D_t(m,E) F_t(m|n_0)}{F_t(m|n_0) - D_t(n_0,E)}, 
\end{equation}
and following the same steps as in the previous sections, we get
\begin{equation}
 \alpha = \lim_{t \to 0} \frac{2~D_t(m,E) F_t(m|n_0)}{D_t(n_0,E)}.
    \label{alpha_SI_cont}
\end{equation}
This completes our derivation of the inference relations for $\alpha$ and $\beta$ for the case of continuous Markov processes. It is to be noted that though Eq.~\eqref{short_threshold} was derived for the detection of $X_m(t)$ above $m$, we expect it to be valid at short-times for detection in any finite sized interval $[m,m+\Delta]$ as well. The reason is that at short times the probability that the process $X_m(t)$ crosses above $m + \Delta$ is negligible. Thus, finite-size effects do not play a role at this level. Clearly, the threshold crossing problem is recovered in the $\Delta \to \infty$ limit.


\end{document}